# Hidden Breathing Kagome Topology in Hexagonal Transition Metal Dichalcogenides

Jun Jung[1] and Yong-Hyun Kim[1,2,*]

[1]*Department of Physics, Korea Advanced Institute of Science and Technology (KAIST), Daejeon 34141, Republic of Korea*

[2]*Graduate School of Nanoscience and Technology, KAIST, Daejeon 34141, Republic of Korea*

*Correspondence to: yong.hyun.kim@kaist.ac.kr

**A Kagome lattice, formed by triangles of two different directions, is known to have many emergent quantum phenomena. Under the breathing anisotropy of bond strengths, this lattice can become a higher-order topological insulator (HOTI), which hosts topologically protected corner states[1]. Experimental realizations of HOTI on breathing Kagome lattices have been reported for various artificial systems[2–7], but not for simple natural materials with an electronic breathing Kagome lattice. Here we prove that a breathing Kagome lattice and HOTI are hidden inside the electronic structure of hexagonal transition metal dichalcogenides (h-TMD). Due to the trigonal prismatic symmetry, $sp^2$-like hybrid $d$-orbitals create an electronic Kagome lattice with anisotropic inter-site and on-site hopping interactions. We demonstrate that HOTI h-TMD triangular nanoflakes host topologically protected corner states, which could be quantum-mechanically entangled with triple degeneracy. Because h-TMDs are easily synthesizable and stable at ambient conditions, our findings open new avenue for quantum physics based on simple condensed matter systems.**

Quantum materials have attracted great interest because of many possible noble applications that are arising from quantum nature, including high-performance quantum computing, lossless energy transport, or highly efficient data storage. The well-known examples of quantum materials are superconductors and quantum hall insulators. These exotic properties originate from large-scale quantum coherence throughout the material and generally emerge at extreme temperature, pressure, or high magnetic field conditions to eliminate all possible fluctuations. In contrast, topological insulators (TI) can have dissipation-less quantum states localized on their boundaries even under moderate conditions. The boundary states are protected from external perturbations by the bulk topology. The Su-Schrieffer-Heeger (SSH) model is a famous example of TI as the one-dimensional (1D) chain with alternating bond strengths[8]. The system becomes topological in the case of stronger inter-cell hopping than intra-cell one. The nontrivial topology generates topologically protected boundary states at the zero-dimensional (0D) endpoints. Higher-order topological insulators (HOTI) have been proposed as a new type of materials, which generalizes the idea of TI[9,10]. For example, a two-dimensional (2D) second-order TI (SOTI) hosts topologically protected corner states at its 0D corners. Naturally, the edge of 2D SOTI would be a conventional TI. Only a few 2D materials have been proposed as HOTI[11–17].

Recently, a breathing Kagome lattice has been proposed as another candidate for HOTI[1]. A normal Kagome lattice is composed of corner-sharing two identical triangles that point in the opposite direction. An electronic tight-binding (TB) model of the normal Kagome lattice shows that it exhibits graphene-like Dirac bands due to the symmetry between the two identical triangles. A bandgap is opened only when the bond strength of two triangles differs, called a breathing Kagome lattice (BKL). Two types of insulators, trivial or non-trivial, are expected depending on the strengths of inter-cell and intra-cell hoppings. A non-trivial HOTI is achieved with stronger inter-cell hopping, analogous to the 1D SSH model. A variety of artificial systems have been experimentally realized with the BKL including acoustic lattices[2,3], CO molecular arrays[4], photonic crystals[5,6], and coupled-resonators[7]. No simple natural material with electronic BKL has been proposed or discovered.

In this study, we propose hexagonal transition metal dichalcogenides (h-TMDs) as the first natural HOTI materials with $d$-orbital electronic BKL. Monolayer h-TMDs are 2D semiconductors with a direct bandgap[18] and known for having a valley pseudospin at $\pm$K points[19,20]. Many have reported that the $d$-orbital character at the valley is inverted, different from what crystal field split theory predicts[21–23]. Such a band inversion is an indicator of non-trivial topology, but no in-depth discussion has been reported. Here we discovered the topological nature of h-TMDs with the hidden BKL originated from strong inter-site $d$-orbital hybridization[24].

As an archetype example of h-TMDs, we construct a minimal three-band TB model of $d$ orbitals for hexagonal $MoS_2$[25]. We start from the five-fold degenerate atomic $d$-orbitals of a Mo atom, and the trigonal prismatic crystal field owing to the neighboring S atoms splits the $d$-orbitals into three classes as $A_1'\{d_{z^2}\}$, $E'\{d_{xy}, d_{x^2-y^2}\}$, and $E''\{d_{xz}, d_{yz}\}$, as shown in Fig. 1(a)[26]. The separation of $A_1'$ and $E'$ orbitals could be characterized with the crystal field split $\Delta_{CF}$. Density functional theory (DFT) band structure analyses show that the band edge states are mostly composed of $A_1'$ and $E'$ orbitals of Mo atoms with a minimal contribution of S atoms. Moreover, the $A_1'$ and $E'$ orbitals are even with respect to the horizontal mirror symmetry ($\sigma_h$) along the $xy$-plane, whereas the $E''$ orbitals are odd. Then, we can think of a triangular lattice only of Mo atoms, as shown in Fig. 1(b), with three $d$-orbitals on each site to represent the band edge states of h-MoS$_2$. We derive TB parameters for the three $d$-orbitals as a basis of the Mo-only triangular lattice using the maximally-localized Wannier function (MLWF) interpolation[27] (see supplementary information for detailed Wannier orbitals in S1 and TB Hamiltonian in S2.1). The three $d$-orbitals TB model well reproduces the band edge states from DFT calculations, as shown in Fig. 1(c,d). Note that the TB bands correctly capture the band inversion of $d$-orbitals at the K. Despite the usefulness of the TB model, the resulting TB parameters in S2.1 seems to be featureless without showing any governing rule.

To analyze better the TB parameters, we adopt the concepts of $sp^2$ and $sd^2$ hybrid orbitals [28,29] under the $C_3$ rotational symmetry to the three $d$-orbitals basis. Consequently, the $A_1'$ and $E'$ orbitals can be transformed to new hybrid $d$-orbitals $\{h_i\}$, defined by the following transformation rule,

$$\begin{pmatrix} h_1 \\ h_2 \\ h_3 \end{pmatrix} = \begin{pmatrix} 1/\sqrt{3} & 0 & \sqrt{2/3} \\ 1/\sqrt{3} & 1/\sqrt{2} & -1/\sqrt{6} \\ 1/\sqrt{3} & -1/\sqrt{2} & -1/\sqrt{6} \end{pmatrix} \begin{pmatrix} d_{z^2} \\ d_{xy} \\ d_{x^2-y^2} \end{pmatrix}. \qquad (1)$$

The wavefunction of $h_i$ is illustrated in Fig. 2(a). The hybrid orbital is shaped as an in-plane or "lie-down" $d_{z^2}$ orbital with the $z$-axis parallel to the $xy$-plane. The three hybrid orbitals $h_i$ have the identical shape and can be rotated 120 degrees to be transformed to each other, as shown in S1. The in-plane $d_{z^2}$ orbital trimer can effectively describe the electronic structure of h-MoS$_2$, similar to the $p_z$ orbital model for graphene. Under the Cartesian and hybrid bases, the nearest neighbor hopping matrix along $\mathbf{R}_1$ in Fig. 1(b) takes the following form,

$$\begin{pmatrix} -0.19 & +0.35 & +0.41 \\ -0.35 & +0.29 & +0.31 \\ +0.41 & -0.31 & -0.08 \end{pmatrix}_{Cartesian} \rightarrow \begin{pmatrix} +0.27 & +0.02 & +0.10 \\ +0.10 & -0.12 & -0.89 \\ +0.02 & +0.05 & -0.12 \end{pmatrix}_{Hybrid} \quad \text{(in eV)}. \quad (2)$$

The negative hopping amplitude represents a bonding between two orbitals. The featureless TB Hamiltonian is now transformed to a TB matrix with a single dominant hopping with amplitude $t_1 = 0.89$ eV. Obviously, the single dominant hopping is associated with the largest electron lobe overlap between two hybrid $d$-orbitals that share the same hexagonal ring, as shown in the inset of Fig. 2(b) (see also SI for all possible direct hybrid $d-d$ couplings with diagrammatic representations in S2.2). The absolute magnitudes of hybrid TB matrix elements are displayed in Fig. 2(b) up to the third nearest neighbors. Clearly, the electronic structure of h-MoS$_2$ is dominated by two major terms; One is the single dominant nearest-neighbor hopping $t_1$, and the other is the on-site crystal field split $\Delta_{CF} = 0.79$ eV in Fig. 1(a). Note that the computed DFT bandgap (1.67 eV) is two times larger than $\Delta_{CF}$. This implies that the bandgap of h-MoS$_2$ has nothing to do with the crystal field split $\Delta_{CF}$, but could be associated with hybridization between Mo $d$-orbitals[24]. More precisely the bandgap is originated from the inter-site hopping $t_1$ or the inter-site hybridization between in-plane hybrid $d$-orbitals, as we will show later.

With help of the hybrid transformation, we discover a hidden BKL in h-MoS$_2$ that preserves its essential electrical structure. We consider only two major parameters, $t_1$ and $\Delta_{CF}$, in the simplified TB model with three $d$-orbitals on a single atomic site. The on-site (3×3) matrix has non-zero diagonal (on-site) elements associated with the crystal field split and zeros for off-diagonal (hopping) terms under the conventional Cartesian basis, whereas it has only non-zero off-diagonal (hopping) terms of $-\frac{1}{3}\Delta_{CF}$ under the new hybrid basis, as follows

$$\begin{pmatrix} -\frac{2}{3}\Delta_{CF} & 0 & 0 \\ 0 & +\frac{1}{3}\Delta_{CF} & 0 \\ 0 & 0 & +\frac{1}{3}\Delta_{CF} \end{pmatrix}_{\text{Cartesian}} \rightarrow \begin{pmatrix} 0 & -\frac{1}{3}\Delta_{CF} & -\frac{1}{3}\Delta_{CF} \\ -\frac{1}{3}\Delta_{CF} & 0 & -\frac{1}{3}\Delta_{CF} \\ -\frac{1}{3}\Delta_{CF} & -\frac{1}{3}\Delta_{CF} & 0 \end{pmatrix}_{\text{Hybrid}}. \quad (3)$$

Therefore, the latter is considered as the "on-site" hopping interaction between three hybrid orbitals. Suppose that each hybrid orbital on a single site forms a different lattice site, as shown in Fig. 2(c), a Kagome lattice emerges with two possible hopping channels, i.e., inter-site hopping $t_1$ and on-site hopping $t_2$, as bond lines. The inter-site $t_1$-hopping represented by solid lines forms downward triangles inside the hexagonal rings of h-MoS$_2$. The on-site hopping $t_2 = \Delta_{CF}/3 = 0.26$ eV, represented by dashed lines, forms upward triangles inside green Mo atoms. As the inter-site hopping is stronger than the on-site one ($t_2/t_1 = 0.29$), h-MoS$_2$ has a hidden BKL with a topological phase[1].

The nontrivial higher-order topology of a BKL can be captured by the position of the Wannier center (WC) inside the unit cell[30]. The WC quantifies the center of the electron; Hence its macroscopic equivalent quantity is the electron contribution to the polarization. On a trivial insulator, the WC is at the atomic site as electrons are confined on atoms. In contrast, the WC position in the nontrivial phase is on the bonding site rather than the atomic site. Because of the $C_3$ rotation and the mirror symmetry along the $yz$-plane ($\sigma_v$) in a BKL, the position of the WC inside the unit cell should be quantized and can thus be used as a topological index[31,32]. This type of TI, distinguished by the WC position, is called an obstructed atomic limit insulator[30].

Here we directly compute the position of the WC for h-MoS$_2$ by integrating the Berry connection throughout the Brillouin zone[33]. The position of the WC would be at the center of either an upward or a downward triangle in a BKL[1–3,32]. Our computation results show that the WC of h-MoS$_2$ is located at the center of the hexagonal ring (the downward triangle), but not at the atomic site, as marked with the red star in Fig. 2(c)[1,32]. This result is also explained with the valence band wavefunction of the BKL model. As $t_1 > t_2 = \Delta_{CF}/3$, we could assume zero $\Delta_{CF}$ as extreme and only consider a BKL with non-zero inter-site hopping $t_1$. Then, the valence band wavefunction is a trimer bonding orbital, of which center is located at the solid downward triangle in Fig. 2(c) (see also SI for single-band Wannierized VBM in S3). Therefore, h-MoS$_2$ with a hidden BKL can be classified as an obstructed atomic limit insulator or HOTI.

Our BKL model reveals that the band inversion at the valley in h-MoS$_2$ is actually a sign of a HOTI. Under the $C_3$ symmetry, the WC position is expressed by the ratio of the expectation values of the $C_3$ rotation on the valence band at the symmetric points Γ and K of the Brillouin zone[2,35]. The valence band at the Γ point has the $A_1'\{d_{z^2}\}$ character, and the expectation value is unity. In contrast, the valence band is inverted at the K point by lowering one of the $E'\{d_{xy}, d_{x^2-y^2}\}$ orbitals. The inverted valence band state is the eigenstate of orbital angular momentum $L_z$[36]. This results in a non-zero polarization, equivalent to a nontrivial WC position.

The BKL can be either topological or trivial depending on the relative strength of anisotropic hoppings, $t_1$ and $t_2$. We could generated two sets of the hopping parameters that produce the exactly same eigenvalues, but different topology; for examples, we use ($t_1 = 1$, $t_2 = 1/3$) and ($t_1 = 1/3$, $t_2 = 1$) for band diagrams in Fig. 2(d,e), respectively. The topological phase is achieved with a stronger inter-site hopping than the onsite hopping in the BKL, i.e., $t_1 > t_2 = \Delta_{CF}/3$, whereas the trivial phase is obtained through a stronger on-site hopping. The orbital character is inverted at the K point in the topological phase, as shown in Fig. 2(d), but not in the trivial phase due to the dominating crystal field. Therefore, we prove that the bandgap and band inversion of h-MoS$_2$ are mainly originated from the inter-site hopping, not the crystal field splitting.

The lie-down $d_{z^2}$ hybrid orbitals can be fundamentally observed on edges and grain boundaries with broken bonds or symmetry. To demonstrate this, we consider a h-MoS$_2$ zigzag nanoribbon with two types of zigzag edges, namely, Mo-terminated (Mo-edge) and S-terminated (S-edge). As shown in Fig. 3(a,b), both the edge states are noticeably represented by the hybrid orbitals and form metallic bands inside the bandgap. Figure 3(c,d) are schematic BKL diagrams for Mo-edge and S-edge, well corresponding to the DFT-based wavefunctions in Fig. 3(a). The Mo-edge in Fig. 3(c) is equivalent to an SSH chain of two hybrid orbitals with alternating hopping strengths. Therefore, the Mo-edge itself is topological due to the stronger inter-site hopping. The S-edge corresponds to dangling-bond states of a single hybrid orbital pointing outward, resembling $p_z$ dangling-bond states at the graphene nanoribbon[37,38].

Based on the non-trivial higher-order bulk topology, h-MoS$_2$ is also expected to host topologically protected corner states. To verify this, we perform DFT calculations to obtain electronic structures of a triangular nanoflake with Mo-edges. The triple-degenerated corner states are spatially localized at three corners of the flake, as shown in Fig. 4(a), in form of a lone hybrid orbital. Energetically, the three corner states are separated from the bulk and edge states, as shown in Fig. 4(b). The three corner states could be quantum-mechanically entangled when partially occupied in spite of their large separation.

It has been reported that the generalized chiral symmetry pins these corner states at the zero-energy of the system[2,4]. Strictly in h-MoS$_2$, the non-zero inter-site hoppings between the same orbitals (i.e. diagonal elements at the hopping matrix) violate the generalized chiral symmetry, slightly away from the exact BKL. Non-Kagome hopping terms have been gradually turned on in the full TB model. The eigenvalue diagram in Fig. 4(c) shows that the corner states of real material are adiabatically connected to the topologically protected corner states with no level crossing with other states. As a result, the triangular nanoflake of h-MoS$_2$ should have topological corners.

This higher-order topology is indeed a general property of group 6 h-TMDs (MX$_2$, M = Mo, W & X = S, Se, Te). Group 6 transition metals are selected not only because h-TMDs are insulators, but also because two valence electrons at the $+4$ oxidation states of transition metal achieve the $1/3$ electron filling conditions for the hidden BKL. For all cases, the three-band TB model well describes the DFT electronic structure (see S4 in SI). Fig. 4(d) shows absolute magnitudes of TB parameters of MX$_2$ using the hybrid orbital basis. The single $t_1$-hopping of a hidden BKL dominates the electronic structure for all h-TMD cases. Therefore, all group 6 h-TMDs are HOTIs, established by $d$-orbital structures in trigonal prismatic geometry, and their triangular flakes could host topologically protected corner states (see S4 in SI). Similarly, we expect that other groups h-TMD would also possess hidden BKLs with being different only in the electron filling.

In summary, we theoretically predict that h-TMDs are HOTIs. Due to a trigonal prismatic atomic structure, $sp^2$-like hybrid $d$-orbitals form an appropriate basis set on these materials. We have found a single dominant nearest-neighbor hopping parameter between hexagon-sharing hybrid orbitals. The minimal hopping model reveal a hidden BKL, naturally leading to HOTI. The nontrivial higher-order topology is characterized by the band inversion at the K point. As a consequence of HOTI, h-TMDs have topological edges and hence topologically protected corners. Because our model provides a microscopic orbital structure of h-TMDs, we anticipate that it would provide fundamental understandings of other properties of h-TMDs. Furthermore, higher-order topology has the potential to make h-TMD as a stable platform for quantum applications including quantum computing.

**Methods**

First-principles DFT calculations were performed using the Vienna ab-initio simulation package (VASP.6.2) with PAW pseudopotentials in the VASP database[42,43]. Electronic structures are obtained under Perdew-Burke-Ernzerhof (PBE) exchange correlation functional[44]. Additional tests with the hybrid functional by Heyd, Scuseria, and Ernzerhof (HSE06[45]) and PBE including spin-orbit coupling give similar results (see S5). Throughout our discussions, we consider the spinless model. We found that the band split induced by spin-orbit coupling (SOC) is smaller than major parameters in our model. Therefore, the main discussion is unaffected by the addition of SOC (see S5). A plane-wave basis cutoff is chosen as 400 eV and $24 \times 24 \times 1$ k-points are sampled. TB parameters have been obtained by maximally localized Wannier functions (MLWF) interpolation[46–48] using Wannier90[49]. Both conventional and hybrid Wannier bases are illustrated in S1. For nanoribbon and nanoflake calculations, edges are passivated by hydrogen atoms. Only hydrogen atoms are relaxed while others are kept fixed (see S5 for the structure relaxation effect). Atomic structures and wavefunction density profiles are drawn with VESTA[50].

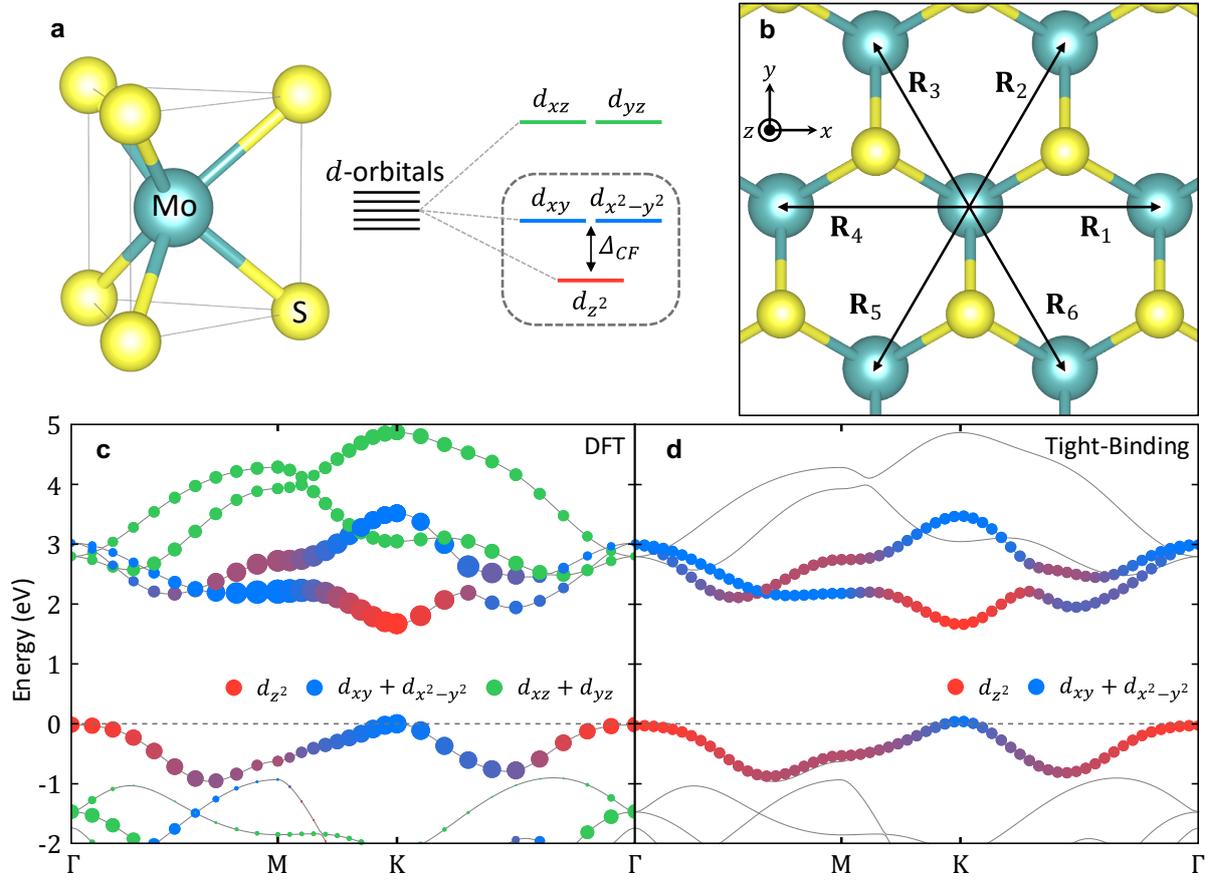

**Fig. 1. Atomic and Electronic Structures of h-MoS$_2$. a,** Trigonal prismatic configuration and corresponding crystal field on $d$-orbitals. The energy difference characterizes the crystal field split strength $\Delta_{CF}$. **b,** The triangular lattice of Mo atoms. Six nearest neighbor vectors $\mathbf{R}_i$ are illustrated. **c,d,** Orbital projected electronic structures of h-MoS$_2$ from (**c**) DFT and (**d**) the three-band TB model. The gray lines represent the reference band eigenvalues from DFT calculations.

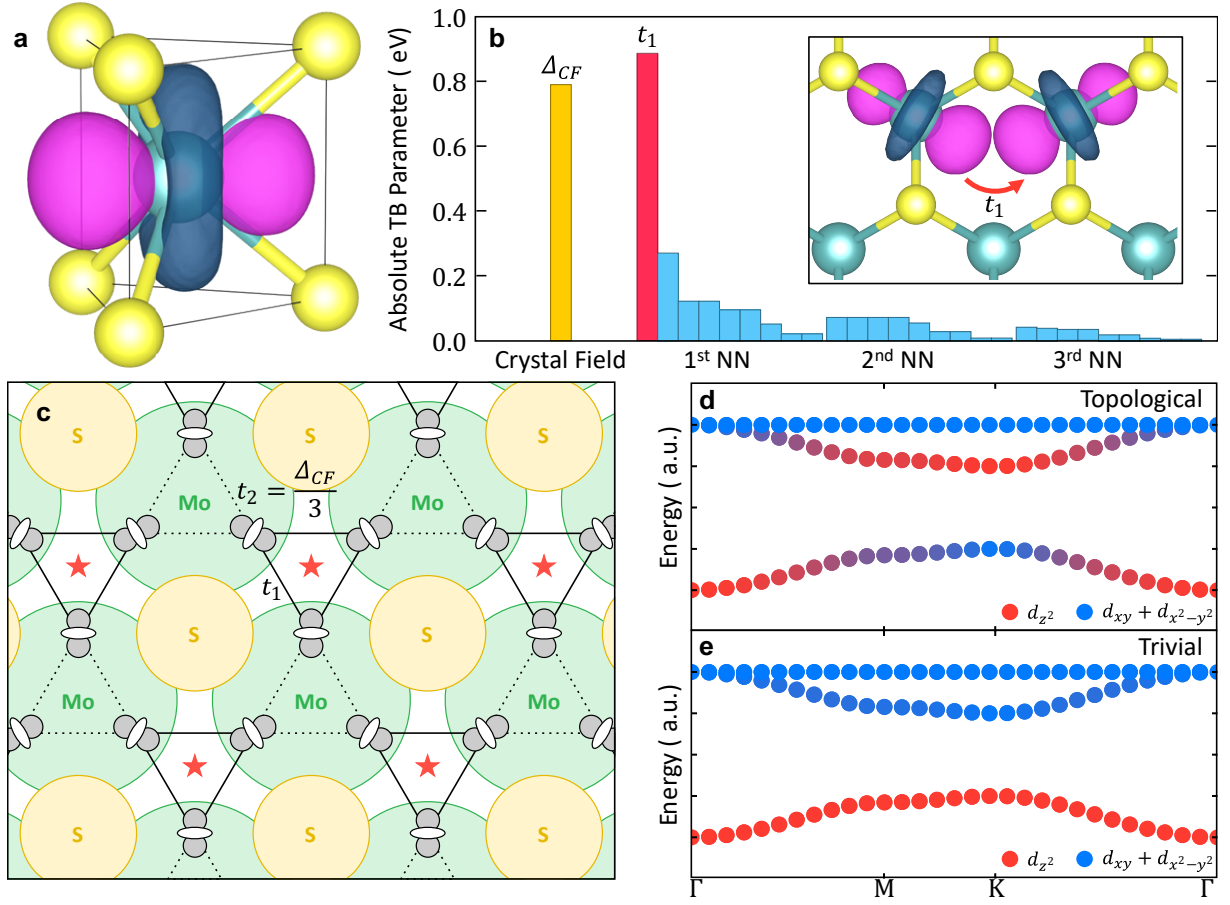

**Fig. 2. Hybrid Orbital and Hidden Breathing Kagome Lattice (BKL). a,** Wavefunction of the hybrid orbital. The hybrid orbital has a shape of an in-plane or "lie-down" $d_{z^2}$ orbital. **b,** Absolute magnitudes of the tight-binding parameters. The NN stands on a nearest neighbor. The inset represents the single-dominant $t_1$-hopping between the 1st nearest neighbors. **c,** A hidden BKL of hybrid $d$-orbitals. The solid lines are the inter-site $t_1$-hopping, and the dashed lines are the on-site $t_2$-hopping originated by the crystal field splitting. **d,e,** Orbital decomposed electronic structures of BKLs. The color code follows Fig. 1(d). A topological phase (**d**) exhibits band inversion, while it is not on a trivial phase (**e**).

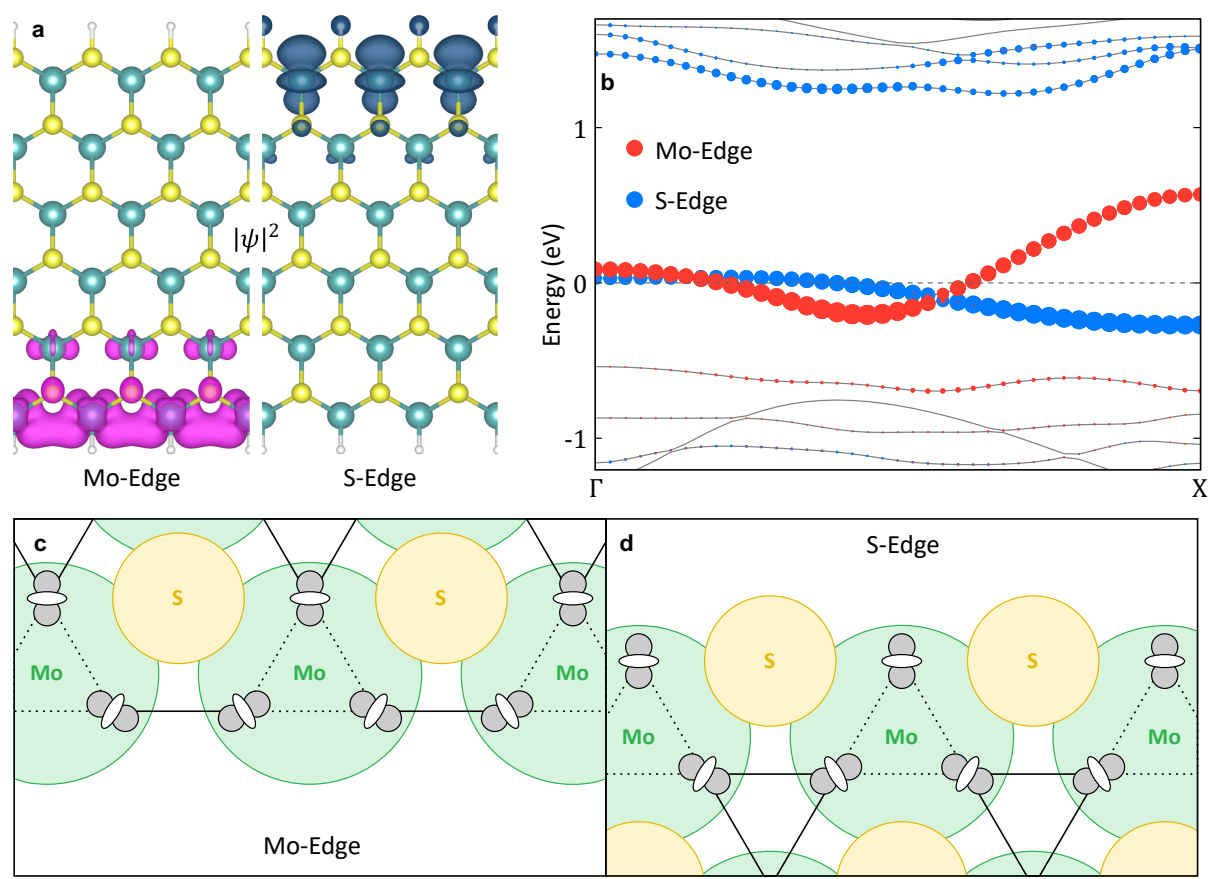

**Fig. 3. 1D Edge States of h-MoS$_2$ Nanoribbon. a**, Wavefunction of edge states on both Mo-edge and S-edge. Mo-edge shows a dimerized wavefunction of hybrid orbitals. S-edge shows a dangling bond like hybrid orbital. **b,** Edge projected band structures of a nanoribbon. Both Mo- and S-edges have edge states inside the bandgap. **c,d,** Schematic orbital diagram of two edges. The Mo-edge (**c**) shows an SSH-model-like alternating chain of hybrid orbitals. On the S-edge (**d**), there exist dangling hybrid orbitals.

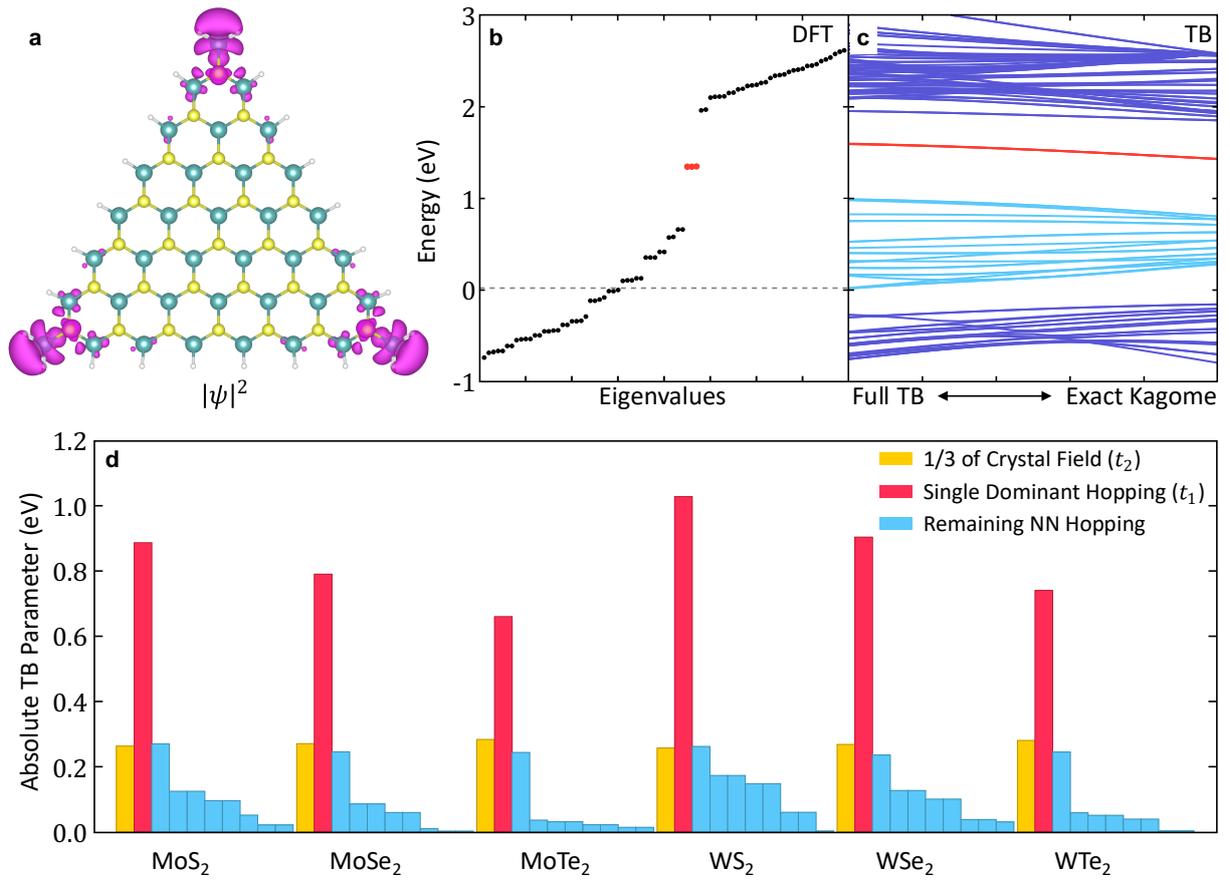

**Fig. 4. Topologically Protected Corner States and Generalization on h-TMDs. a,** Wavefunction of the corner state. It is localized on three corners, and has a shape of a hybrid orbital on each corner. **b,** DFT eigenvalues of a triangular nanoflake. The corner states (red) are located inside the bandgap. **c,** TB eigenvalues of a triangular nanoflake. Hopping parameters are gradually changed from the exact BKL to the full TB model. The corner states of the full TB model are adiabatically connected to the exact BKL model. **d,** Absolute magnitudes of TB parameters under hybrid orbitals for group 6 h-TMDs. They are all HOTIs due to the strongest inter-site $t_1$-hopping terms.

# Supplementary Information: Hidden Breathing Kagome Topology in Hexagonal Transition Metal Dichalcogenides

Jun Jung[1] and Yong-Hyun Kim[1,2,*]

[1]*Department of Physics, Korea Advanced Institute of Science and Technology (KAIST),*

*Daejeon 34141, Republic of Korea*

[2]*Graduate School of Nanoscience and Technology, KAIST, Daejeon 34141, Republic of*

*Korea*

*Correspondence to: yong.hyun.kim@kaist.ac.kr

## S1. Wannier Orbitals

Wannier basis represents antibonding character between Mo $d$ orbitals and S $p$ orbitals. Their real space wavefunctions are drawn below.

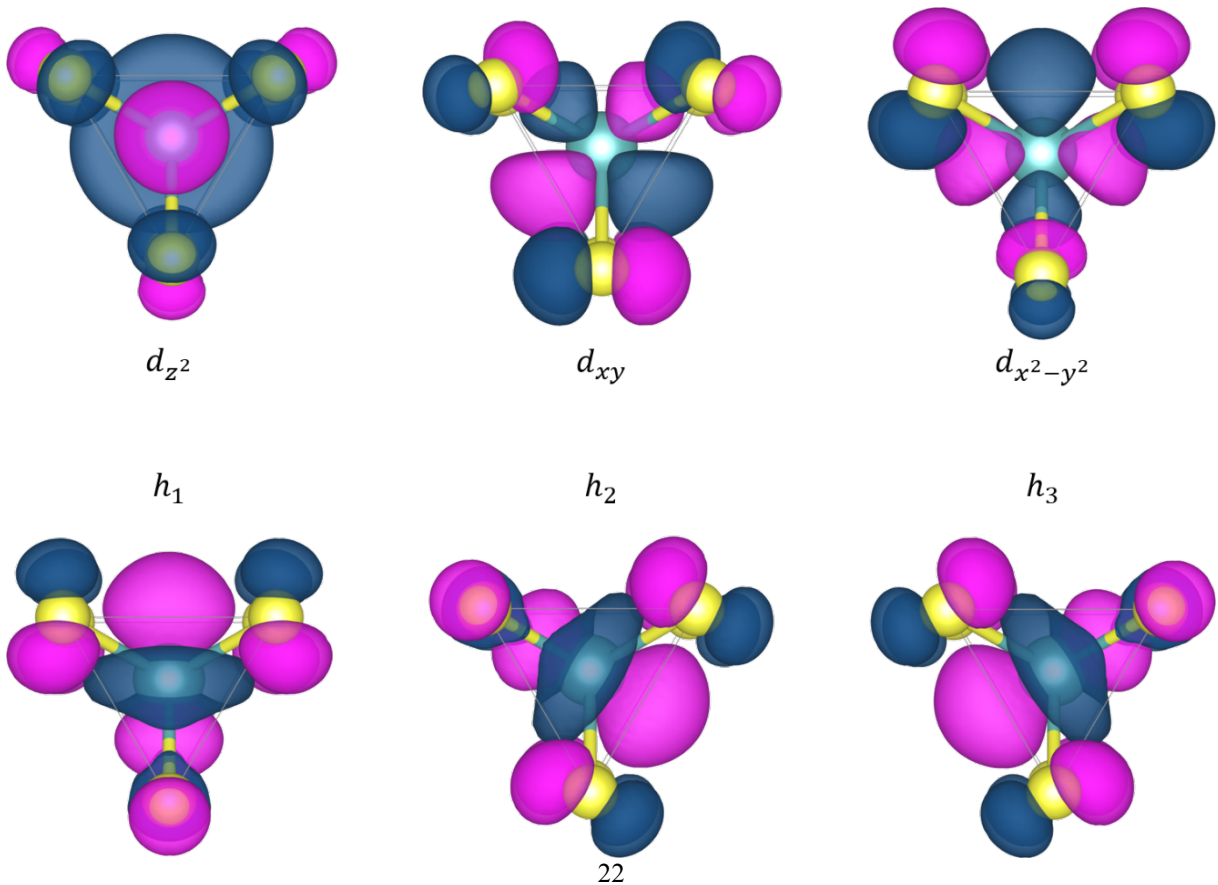

$d_{z^2}$  $d_{xy}$  $d_{x^2-y^2}$

$h_1$  $h_2$  $h_3$

## S2. Tight Binding Model

### S2.1 Hopping Parameters

The tight binding Hamiltonian in momentum space is written as

$$H(\mathbf{k}) = \sum_{\mathbf{R}} t(\mathbf{R}) e^{-i\mathbf{k}\cdot\mathbf{R}}, \qquad (1)$$

where $\mathbf{R}$ stands for every lattice vector. For nearest neighbor vectors $\mathbf{R}_1$ and $\mathbf{R}_2$, the hopping parameters up to the third nearest neighbor hopping have the following form. The subscripts on a matrix are C for a conventional basis and H for hybrid basis.

$$t(\mathbf{R}_1) = \begin{pmatrix} -0.19 & +0.35 & +0.41 \\ -0.35 & +0.29 & +0.31 \\ +0.41 & -0.31 & -0.08 \end{pmatrix}_C \rightarrow \begin{pmatrix} +0.27 & +0.02 & +0.10 \\ +0.10 & -0.12 & -0.89 \\ +0.02 & +0.05 & -0.12 \end{pmatrix}_H \qquad (2)$$

$$t(\mathbf{R}_1 + \mathbf{R}_2) = \begin{pmatrix} +0.05 & -0.08 & +0.05 \\ +0.02 & +0.10 & -0.03 \\ -0.01 & -0.03 & +0.06 \end{pmatrix}_C \rightarrow \begin{pmatrix} +0.07 & -0.07 & +0.03 \\ +0.01 & +0.06 & +0.01 \\ +0.03 & -0.07 & +0.07 \end{pmatrix}_H \qquad (3)$$

$$t(2\mathbf{R}_1) = \begin{pmatrix} -0.02 & +0.01 & -0.00 \\ -0.01 & +0.08 & +0.31 \\ -0.00 & +0.01 & +0.02 \end{pmatrix}_C \rightarrow \begin{pmatrix} +0.01 & -0.00 & -0.02 \\ -0.02 & +0.04 & -0.04 \\ -0.00 & -0.04 & +0.04 \end{pmatrix}_H \qquad (4)$$

The remaining hopping matrices can be obtained by applying $C_3$ rotation, mirror, and reciprocal relation $t(-\mathbf{R}) = t(\mathbf{R})^T$.

## S2.2 Diagrammatic Expression

Inter-site hopping between hybrid orbitals is drawn below.

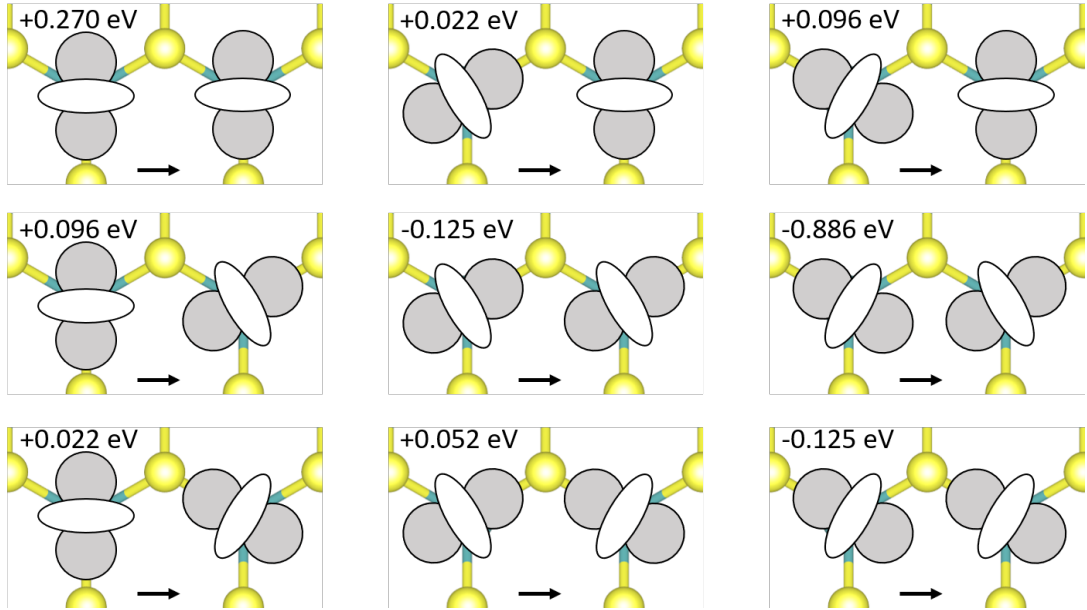

The orbital overlap $\langle \psi_j(\mathbf{r} + \mathbf{R}) | \psi_i(\mathbf{r}) \rangle$ below is calculated between nearest neighbors. Atomic $d$-orbitals are generated by 13-band MLWF interpolation. Due to the electrostatic repulsion of surrounding anions, orbital weight is larger along the hexagonal ring.

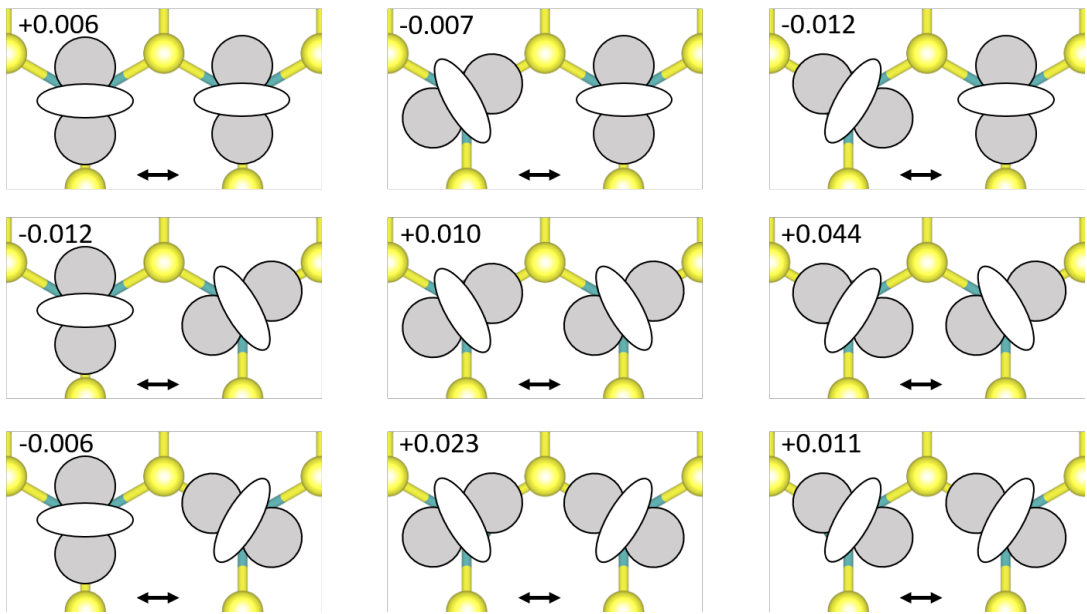

## S3. Single-Band VBM Wannierization

Wannier interpolation TB model can be constructed only on the VBM band. The resulting Wannier orbital exhibits a molecule orbital-like wavefunction, centered at the hexagonal ring. This orbital can be considered as a trimer bonding state of three hybrid orbitals, sharing the same hexagonal ring.

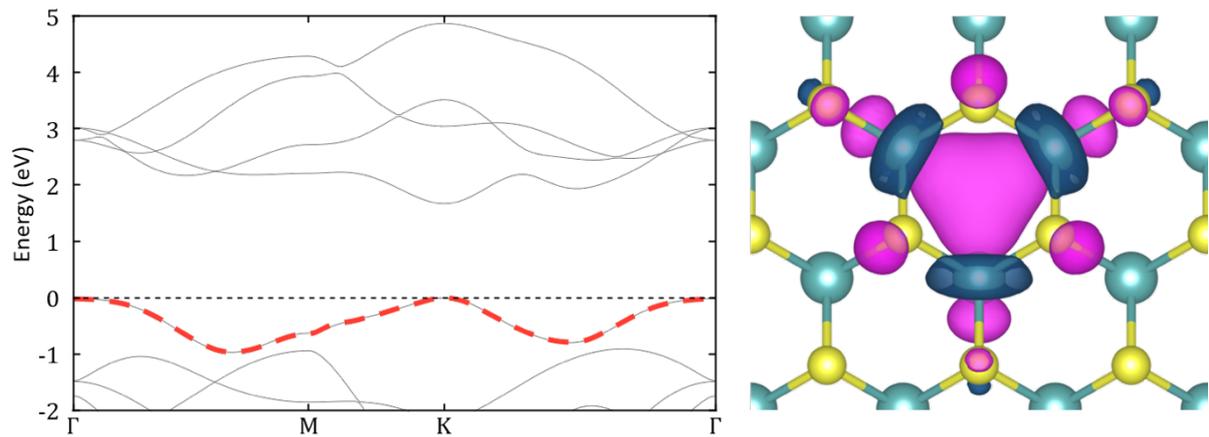

## S4. Generalization on all h-TMDs

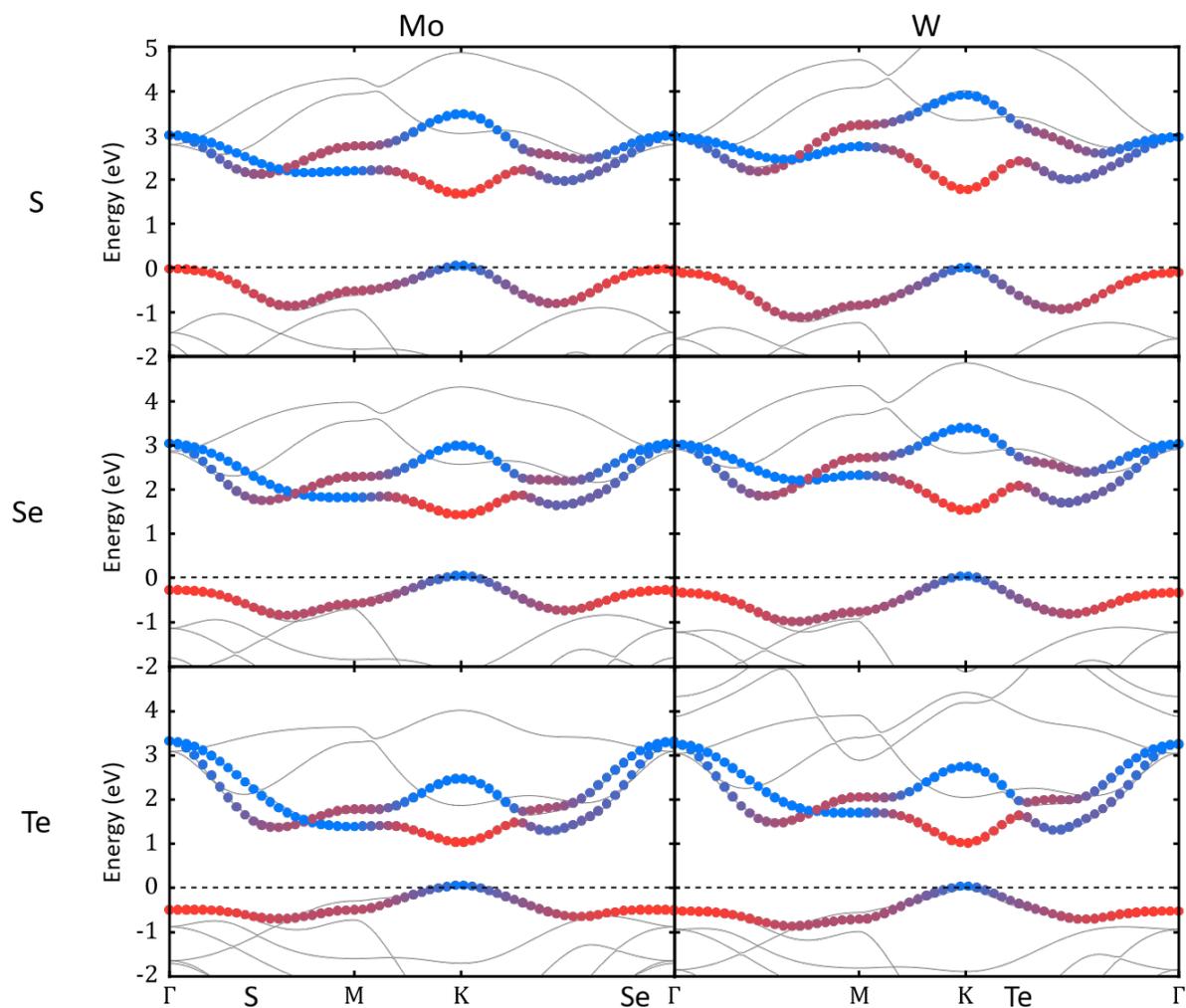

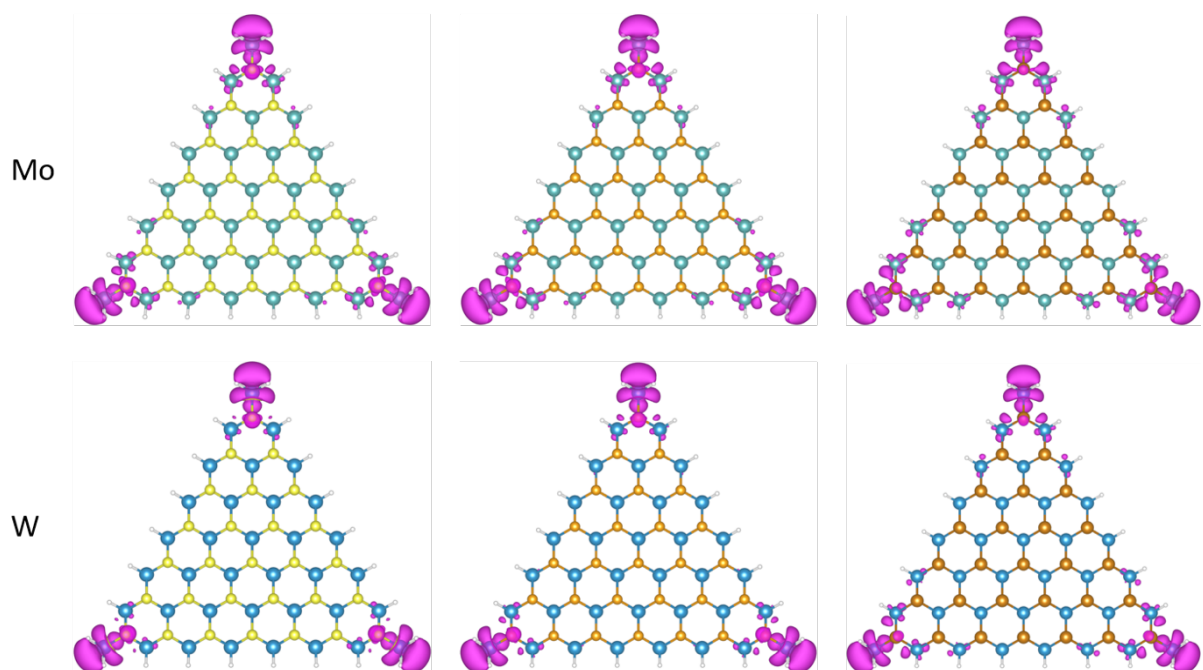

**S5. SOC and HSE**

The topologically protected corner states remain stable under the consideration of SOC. Under a hybrid functional test without SOC, the system still exhibits the nontrivial higher-order topology.

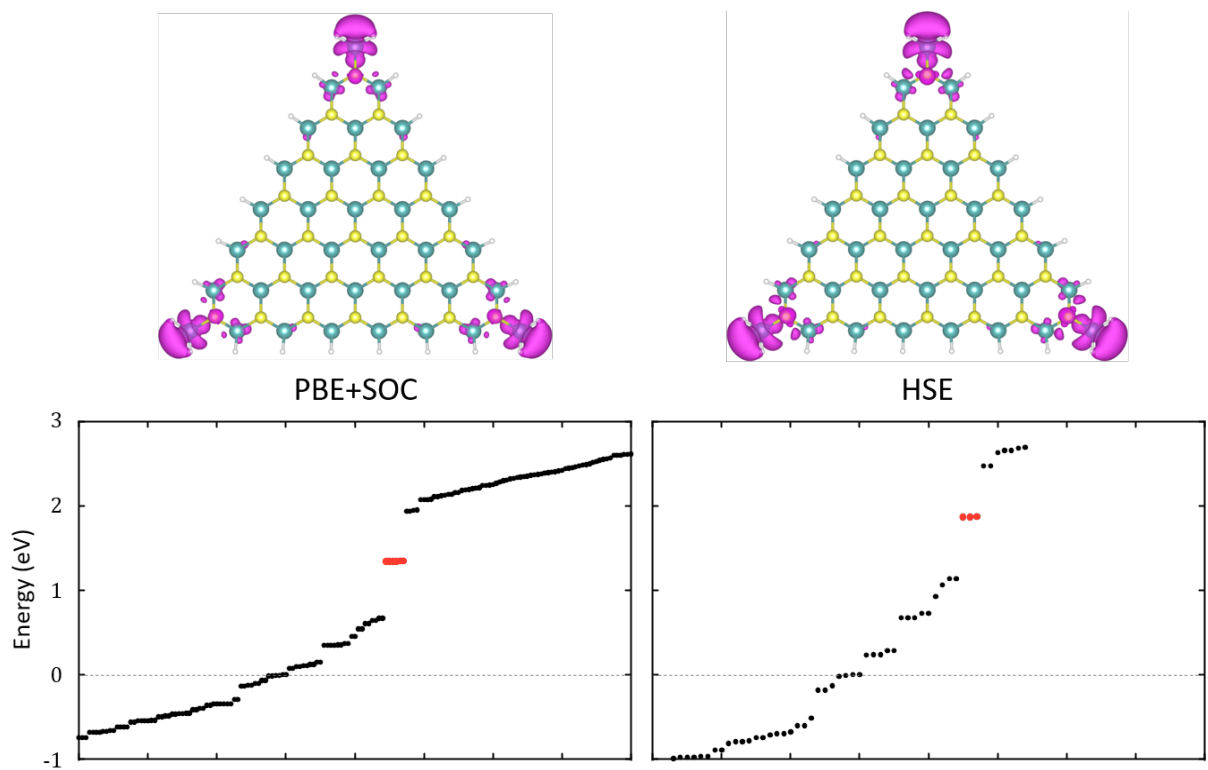

## S6. Structure Relaxation Effect

In the main text, Mo and S atoms keep the original structure and only the passivating hydrogen atoms are relaxed. Eigenvalues are plotted as structure has fully relaxed. Some of edge states become nearly degenerate to the corner states.

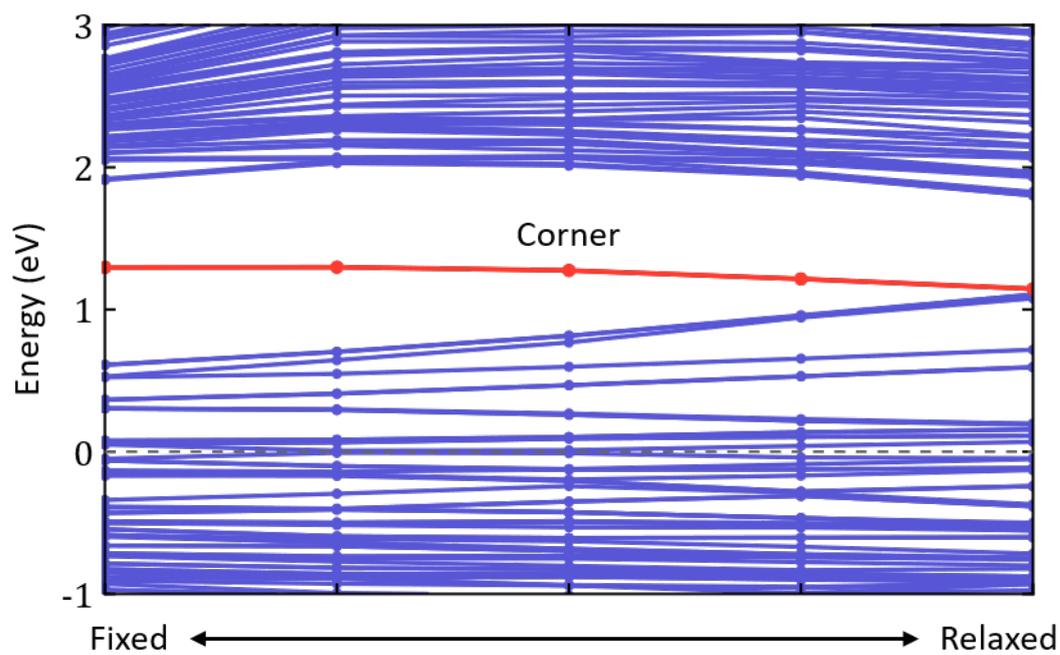

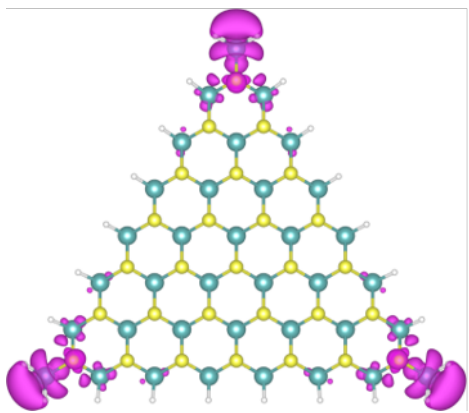
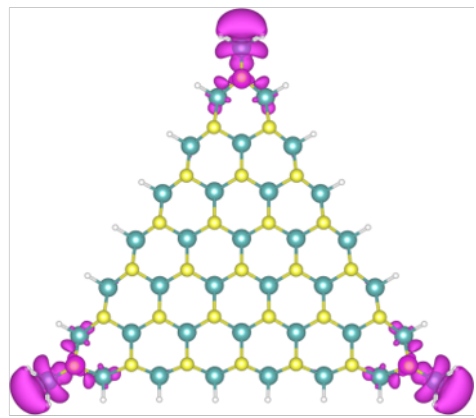